\documentclass[aps,prb,twocolumn,showpacs]{revtex4}

\usepackage{graphicx}  

\begin{document}

\title{Prediction of huge X-ray Faraday rotation at the Gd N$_{4,5}$ threshold}

\author{J.E. Prieto, F. Heigl, O. Krupin, G. Kaindl and K. Starke} 
\affiliation{Institut f\"ur Experimentalphysik, Freie Universit\"at Berlin, 
Arnimallee 14, D-14195 Berlin, Germany}

\date{\today}

\begin{abstract}

X-ray absorption spectra in a wide energy range around the 4d--4f excitation 
threshold of Gd were recorded by total electron yield from in-plane 
magnetized Gd metal films. 
Matching the experimental spectra to tabulated absorption data 
reveals unprecedented short light absorption lengths 
down to 3~nm. The associated real parts of the refractive index 
for circularly polarized light propagating parallel or antiparallel 
to the Gd magnetization, determined through the Kramers-Kronig 
transformation, correspond to a magneto-optical Faraday rotation of 
${0.7^{\circ}}$ per atomic layer.
This finding shall allow the study of magnetic structure 
and magnetization dynamics of lanthanide elements in
nanosize systems and dilute alloys.

\end{abstract}

\pacs{75.30.-m, 75.70.-i, 78.20.Ls, 78.70.Dm}
\keywords{}

\maketitle

Resonance enhancements of X-ray magnetic scattering cross sections
at inner-shell absorption edges\cite{gibbs88,imp89}
have been used for years to investigate the magnetic structure 
of lanthanide\cite{mgb99} and actinide\cite{mll99} systems
in the {\em hard} X-ray regime (above ${\rm 2~keV}$).
The experimental demonstration of large changes in the specularly 
reflected X-ray intensity at the ${\rm Fe~L_{2,3}}$ edge\cite{cck94} 
upon magnetization reversal initiated the ongoing search for 
magneto-optical (MO) effects in the {\em soft} X-ray regime, 
\cite{tonnerre95,hks95,cic98,shp98,kok00,schuessl01,mog01} 
as well as their application to element-specific studies of heteromagnetic 
systems.\cite{tsb98,icf99,wbh99,kko00,hkt00,ggj01}

Yet, in analyzing soft X-ray MO signals from thin films and 
multilayer systems with thicknesses comparable to the X-ray wavelength, 
previous
investigations have shown\cite{cck94,tsb98,ggj01} that 
a comparison with model calculations\cite{sts00} 
of the reflected specular intensity (based on the Fresnel equations) 
is needed in order to extract a layer-resolved sample magnetization profile.
Several experimental determinations of soft X-ray MO constants
have been reported\cite{cic98,shp98,kok00} 
for ferromagnetic transition metals 
in the region of the ${\rm L_{2,3}}$ thresholds, 
but none so far for the lanthanide elements,
despite their wide recognition as, e.g., constituents of 
exchange-spring magnets\cite{fjs98} and magnetic recording media\cite{nak99}.
Only recently has it been demonstrated that sizeable MO signals are
obtained from lanthanide elements in the soft X-ray region 
at the ${\rm N_{4,5}}$ thresholds.\cite{shv01}

Here we show that calibrated 
${\rm N_{4,5}}$ absorption spectra from magnetized Gd metal,
recorded with circularly polarized (CP)
light in the energy interval from 110 to 200~eV, 
yield an X-ray absorption coefficient up to three times larger 
than expected. The magnetization-dependent absorption of CP light at the 
Gd ${\rm N_{4,5}}$ giant resonance maximum, described by the imaginary part 
of the refraction index, is accompanied by a huge change
in light propagation speed upon magnetization reversal, a dispersive
effect described by the real part of the refraction index, 
implying a Faraday rotation (FR) of about ${0.7^{\circ}}$ per atomic layer.
Thus, even very small or diluted lanthanide systems are expected to show a 
measurable effect. 

The absorption experiments were performed at the high-resolution UE56 
undulator beamline\cite{UE56} of the Ber\-li\-ner Ele\-ktro\-nen\-spei\-cher\-ring
f\"ur Syn\-chro\-tron\-strah\-lung (BESSY II). The photon energy resolution 
was set to about ${\rm 100~meV}$~(FWHM) which is well below the
intrinsic width of the narrow Gd ${\rm N_{4,5}}$ pre-edge absorption
lines.\cite{sna97} 
The photon energy interval from 110 to 200~eV was scanned 
at slow speed by a synchronized movement
of monochromator and undulator. This synchronization is essential
to properly normalize the absorption spectra and allows one to exploit the
high flux of the undulator beamline of about 
${\rm 10^{14}\,photons/(s\cdot 100~mA\cdot 0.1\%\,bandwidth)}$ 
over a wide energy range.
The degree of circular polarization at this Sasa\-ki-type undulator beamline
is practically ${\rm 100~\%}$.\cite{UE56}

The absorption spectra were recorded in total-electron yield (TEY) mode
using a high-current channeltron.
To suppress the background of secondary electrons from the
chamber walls, both the sample and a retarding grid placed in front of 
the channeltron, were biased with a low-voltage battery.  
For signal stability, high voltage was supplied to the channeltron cathode 
by a 3.2~kV battery box.
The electron-yield current was amplified by an electrometer (set to 3~ms 
integration time for a scan speed of typically 0.1~eV per second).
We used a light incidence angle of $30^{\circ}$ with respect
to the film plane, in order to compromise between a large projection of the CP
light wave-vector onto the in-plane film magnetization and the desired small 
sample reflectivity.\cite{ahs97}

Epitaxial Gd metal films of (10~$\pm$~1)~nm thickness were prepared 
{\em in situ} by vapor deposition in ultra-high vacuum 
(${\rm 3 \times 10^{-11}~mbar}$ base pressure; about ${\rm 4 \times 
10^{-10}~mbar}$ during deposition) on a W(110) single-crystal substrate
(for details of film preparation, see Ref.~\onlinecite{sta00}).
For remanent {\em in-plane} sample magnetization, an external field 
was applied along [1$\bar{1}$0] of the substrate, i.e., parallel to 
the easy magnetization axis of the Gd film, using 
a rotatable electromagnet.\cite{magnet} A compact visible-light 
MO Kerr-effect setup\cite{sed92} was used to routinely check the state 
of remanent magnetization of the Gd films, revealing square-shaped 
hysteresis loops with about ${\rm 100~Oe}$ coercivity.

\begin{figure}
\includegraphics*[width=8.0cm]{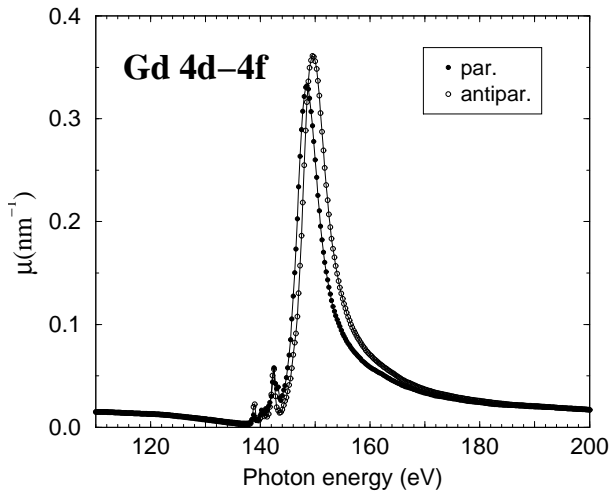}
\caption{\label{mupm} \small Gd ${\rm N_{4,5}}$ absorption spectra 
from remanently magnetized
Gd films at ${T = 30~K}$. CP light was incident at $30^{\circ}$
with respect to the film plane, i.e. mainly parallel (filled symbols) and 
antiparallel (open symbols) to the in-plane sample magnetization.}
\end{figure}

Figure~\ref{mupm} displays experimental absorption spectra in the region of the 
Gd ${\rm N_{4,5}}$ threshold. 
The spectra were corrected for saturation\cite{vt88,nsi99}
assuming 0.3 for the ratio of electron escape depth to minimal X-ray 
absorption length. The photon energy range of the present spectra
is significantly wider than of those measured in previous 
studies\cite{sna97,muto94}, including the wide asymmetric 
flanks of the ${\rm 4d\rightarrow 4f}$ 
giant resonance (Beutler-Fano profile). This allows one to 
calibrate the absorption spectra by matching
both ends to the tabulated absorption coefficient\cite{henke}
at photon energies where the influence of the giant resonance is
expected to be negligible. 
To this end we fixed the absorption coefficients $\mu_\pm$ at the low-energy 
(110~eV) and high-energy sides (200~eV) of the measured spectra to the values
given by the tables of Henke {\em et al.}\cite{henke}, 15.0 and 17.1~$\times
10^{-3}~nm^{-1}$ at 110 and 200~eV, respectively. This then defines the given 
scale of the ordinate in Fig.~\ref{mupm}. 
In this way, the absoption coefficient is obtained with an error bar 
of $\pm$15\% at the maximum, estimated from our experimental precision of
$\pm$1\% at both ends of the photon energy range, where the matching 
to the tabulated data was performed. 

We note that the spectra in Fig.~\ref{mupm} show the same qualitative energy 
dependence as given earlier\cite{sna97},
yet the previous spectra were scaled to the same maximum value for 
both magnetization directions without any correction for saturation 
effects.

\begin{figure}
\includegraphics*[width=8.0cm]{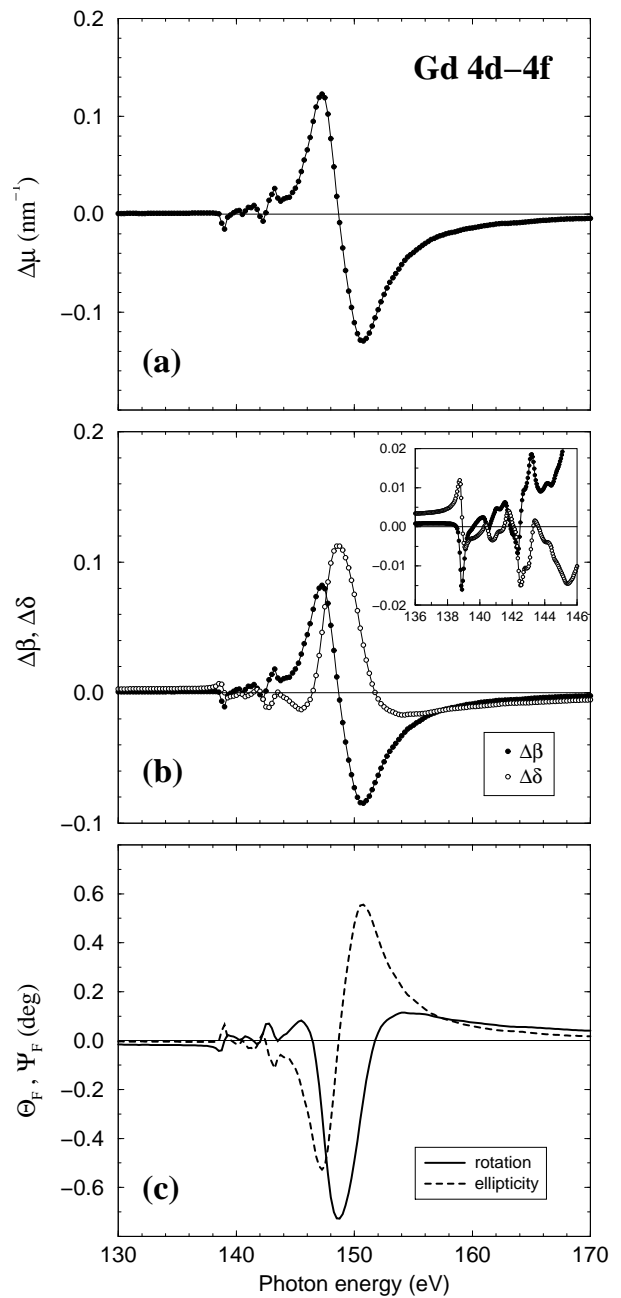}
\caption{\label{delta}  \small (a) Difference spectrum,
${\rm \Delta \mu(\omega) \equiv (\mu_+(\omega) - }$ 
${\rm \mu_-(\omega)) / cos(30^{\circ})}$, of the absorption spectra 
for opposite magnetizations given in Fig.~\ref{mupm}.
(b) Difference ${\rm \Delta \beta}$ of the imaginary parts (filled symbols) 
and the associated difference ${\rm \Delta \delta}$ of the real parts 
(open symbols) obtained through a Kramers-Kronig transformation. 
Inset: pre-edge range, measured (${\rm \Delta \beta}$, filled symbols)
and calculated (${\rm \Delta \delta}$, open symbols) with higher point density.
(c) FR and ellipticity spectra calculated for linearly 
polarized light transmitted normally through a 0.3~nm thick Gd metal film
magnetized perpendicular to the film plane.}
\end{figure}

The comprehensive X-ray data tables of Henke~{\em et al.}\cite{henke} 
contain the lanthanide ${\rm N_{4,5}}$ absorption spectra by 
Zimkina~{\em et al.}\cite{zfg67},
who measured {\em relative} linear X-ray absorption lengths
of nonmagnetized lanthanide samples.
To obtain an absolute absorption length, Henke~{\em et al.}\cite{henke} 
followed Richter~{\em et al.}\cite{rmp89} 
who calibrated their gas-phase photoexcitation data by
matching them to calculated cross sections.
In this way they arrived at a maximum absorption coefficient
at the nonmagnetic Gd ${\rm N_{4,5}}$ peak of ${\rm \mu \approx 0.1~nm^{-1}}$,
corresponding to a linear X-ray absorption length 
${\rm \lambda = 1/\mu \approx 10~nm}$.\cite{henke}
By contrast, the calibrated experimental spectra 
from magnetized Gd in Fig.~\ref{mupm} reveal maximum
values for the absorption coefficient 
of ${\rm 0.33~nm^{-1}}$ and ${\rm 0.36~nm^{-1}}$, for nearly parallel 
and antiparallel orientation of sample magnetization and photon spin, 
respectively.
The corresponding linear absorption lengths are 3.0~nm and 2.8~nm,
with the quoted error of $\pm$15\%;
they are about three times shorter than expected.\cite{henke} 
These soft X-ray absorption lengths are remarkably short, even compared with
visible-light absorption lengths in metals of typically some ${\rm 20~nm}$.

The difference spectrum 
${\rm \Delta \mu(\omega) \equiv (\mu_+(\omega) - \mu_-(\omega)) /}$ 
${\rm  cos(30^{\circ})}$, 
obtained from the experimental absorption spectra 
${\rm \mu_{\pm}(\omega)}$, is displayed in Fig.~\ref{delta}a. 
The factor  ${\rm 1 / cos (30^{\circ})}$ accounts for the 
finite experimental angle between the direcions of light propagation 
and magnetization.
Apart from minor contributions from the weaker pre-edge transitions 
(inset of Fig.~\ref{delta}b)
around ${\rm 140~eV}$, the ${\Delta \mu(\omega)}$ spectrum exhibits an
S-shape behavior, with a zero crossing near ${\rm 149~eV}$.
It originates from the very intense 
${\rm 4d^{10}~4f^{7} [^8S] \rightarrow 4d^{9}~4f^{8} [^8P]}$ transitions 
(dipole-allowed in LS coupling). For {\em parallel} orientation of 
photon spin and sample magnetization (${\rm \Delta M = +1}$ transitions),
the intermediate~$^8P_{5/2}$ state at around ${\rm 148~eV}$
is preferentially populated; for {\em antiparallel} orientation 
(${\rm \Delta M = -1}$ transitions), by contrast, the only allowed 
excitation is into the
higher ${\rm ^8P_{9/2}}$ state at around ${\rm 150~eV}$.\cite{sna97}
The large difference in absorption coefficient for opposite magnetization 
directions (Fig.~\ref{delta}a) corresponds to a difference in the absorptive 
part ${\Delta \beta \equiv \beta_+ - \beta_- = \Delta \mu\,\lambda / (4 \pi)}$
of the refractive index ${n_{\pm} = 1-\delta_{\pm} -i \beta_{\pm}}$. 
As shown in Fig.~\ref{delta}b, ${\Delta \beta }$ changes from
${+ 0.082}$ to ${- 0.085}$ within ${\rm 3.5~eV}$.

From the present data, the associated difference in the real part 
${\Delta \delta \equiv \delta_+ - \delta_-}$, for CP light propagating
(exactly) parallel or antiparallel to the Gd magnetization, 
was derived using the Kramers-Kronig (KK) transformation 
for magnetic systems;\cite{pva99} the result is given in Fig.~\ref{delta}b
by open symbols. The accuracy of this integral transformation depends 
mainly on the spectral range available for integration. 
Within the extended photon energy range of 
110 - 200~eV, the absorption spectra recorded for opposite 
magnetization directions appear to become asymptotically 
equal at both ends of the experimental photon-energy range 
(see Fig.~\ref{mupm}). 
Hence the difference ${\Delta \beta}$ vanishes at the two boundaries, 
so that the result of the KK transformation is not affected by the 
choice of the experimental photon energy range. 
${\Delta \delta}$ peaks right at the zero crossing of 
the absorptive part, where it amounts to ${\Delta \delta \approx 0.11}$ 
(see Fig.~\ref{delta}b).

For future applications, we use the experimental 
difference in the absorptive part, ${\Delta \beta}$, together with
the calculated phase difference, ${\Delta \delta}$, to calculate
the complex FR of 
Gd metal. Here we assume fully oriented 4f magnetic moments
as existing, e.g., in the ferromagnetic phase at low temperatures
($T / T_C \ll 1$). Real and imaginary parts of the FR, 
$\Theta_F$ and $\Psi_F$, respectively, are given by the 
expressions\cite{zvk97}
\begin{eqnarray}
\tan (2\Theta_F) & = & 2\, Re\left[ a\right] / (1  - \left| a\right|^2)\;,
\\
\sin (2\Psi_F) & = & 2\, Im\left[ a\right] / (1  + \left| a\right|^2)\;,
\end{eqnarray}
\noindent
where $d$ is the film thickness, ${a = \tan( \Delta n \omega d / c )}$ 
and ${\Delta n = (n_+ - n_-)/2 }$. The $\Theta_F(\omega)$ and 
$\Psi_F(\omega)$ spectra of Gd metal at the ${\rm N_{4,5}}$ threshold are 
presented in Fig.~\ref{delta}c for linearly polarized (LP) light transmitted 
in normal direction through a 0.3~nm (1~monolayer)
thick Gd metal film magnetized perpendicular to the film plane, either
{\em parallel} or {\em antiparallel} to the light propagation direction. 
The spectra predict a FR of 
$\Theta_F = (0.73\pm0.11)^{\circ}$ per 0.3~nm 
(${\rm (2.4\pm0.4)^{\circ} / nm }$) 
near ${\rm 149~eV}$, right at the zero crossing of the absorption 
difference for CP light in Fig.~\ref{delta}a.
At this photon energy, $\Theta_F$ is accompanied by a vanishing
Faraday ellipticity $\Psi_F$ (cf. Fig.~\ref{delta}c).
To our knowledge this is by far the largest specific FR reported. It is 
9~times larger than the specific rotation maximum at the Fe ${\rm L_3}$ 
threshold\cite{kok00} and some 70~times (50~times) larger than in the 
visible (infrared) region of Fe metal.

With the predicted specific FR at N$_{4,5}$ thresholds, already some
${\rm 10^{15}}$ lanthanide atoms per cm$^{2}$ as in, e.g., a single
atomic layer, a very dilute film, or nanosize particles, should be
sufficient to show a measurable rotation. Note that it is not at all
evident that continuum classical electrodynamics, as used in this
work for the 10~nm thick Gd metal films, will still be appropriate
when approaching atomic dimensions.

The huge FR at ${\rm Gd~N_{4,5}}$ is due to the very large 
electric dipole (E1) transition probability of 
${\rm 4d^{10}~4f^{n}\rightarrow 4d^{9}~4f^{n+1}}$ transitions 
($n=7$ for Gd).
Hence, when applying magnetized Gd films as a method to rotate the plane 
of light polarization at the fixed photon energy of ${\rm 149~eV}$, 
the strong absorption at the ${\rm Gd~N_{4,5}}$ maximum (cf. Fig.~\ref{mupm}) 
leads to a very short penetration length of the order of only 3~nm. 
In order to obtain, e.g., a FR of ${\pm 45^{\circ}}$ for opposite
film magnetizations, one would use a 18.5~nm thick Gd film, with an 
inevitable intensity reduction by a factor of ${\rm 4.5 \times 10^{2}}$.
Despite this substantial loss in intensity, which leads to a transmitted 
flux of about ${\rm 10^{11}\,photons/(s\cdot 100~mA\cdot 0.1\%\,bandwidth)}$ 
at a typical third-generation undulator beamline,
Gd films might well be useful in differential (lock-in technique) 
experiments, where {\em fast switching} of the X-ray polarization plane 
(ns time scale) is required.  One could extend the photon energy range 
of this method to about ${\rm 180~eV}$\cite{henke} by using heavier 
lanthanide elements.

\begin{acknowledgments}
J.E.P. thanks the Alexander-von-Humboldt Stiftung for generous 
support. We gratefully achnowledge the experimental help of Fred Senf 
and Rolf Follath (BESSY), and useful discussions with Jeff Kortright 
and Eric Gullikson (LBNL). This work was financed by the German 
Bundesministerium f\"ur Bildung und Forschung, contract no. 05 KS1 KEC/2.
\end{acknowledgments}

\bibliography{faraday}

\begin{thebibliography}{10}
\expandafter\ifx\csname bibnamefont\endcsname\relax
  \def\bibnamefont#1{#1}\fi
\expandafter\ifx\csname bibfnamefont\endcsname\relax
  \def\bibfnamefont#1{#1}\fi
\expandafter\ifx\csname url\endcsname\relax
  \def\url#1{\texttt{#1}}\fi
\expandafter\ifx\csname urlprefix\endcsname\relax\def\urlprefix{URL }\fi
\expandafter\ifx\csname bibinfo\endcsname\relax \def\bibinfo#1#2{#2}\fi
\expandafter\ifx\csname eprint\endcsname\relax \def\eprint#1{#1}\fi

\bibitem{gibbs88}
\bibinfo{author}{\bibfnamefont{D.}~\bibnamefont{Gibbs}},
  \bibinfo{author}{\bibfnamefont{D.~R.} \bibnamefont{Harshman}},
  \bibinfo{author}{\bibfnamefont{E.~D.} \bibnamefont{Isaacs}},
  \bibinfo{author}{\bibfnamefont{D.~B.} \bibnamefont{McWhan}},
  \bibinfo{author}{\bibfnamefont{D.}~\bibnamefont{Mills}}, \bibnamefont{and}
  \bibinfo{author}{\bibfnamefont{C.}~\bibnamefont{Vettier}},
  \bibinfo{journal}{Phys. Rev. Lett.} \textbf{\bibinfo{volume}{61}},
  \bibinfo{pages}{1241} (\bibinfo{year}{1988}), \bibinfo{note}{who discovered
  x-ray resonant magnetic scattering at the Ho $L_{3}$ absorption threshold.}

\bibitem{imp89}
\bibinfo{author}{\bibfnamefont{E.~D.} \bibnamefont{Isaacs}},
  \bibinfo{author}{\bibfnamefont{D.~B.} \bibnamefont{McWhan}},
  \bibinfo{author}{\bibfnamefont{C.}~\bibnamefont{Peters}},
  \bibinfo{author}{\bibfnamefont{G.~E.} \bibnamefont{Ice}},
  \bibinfo{author}{\bibfnamefont{D.~P.} \bibnamefont{Siddons}},
  \bibinfo{author}{\bibfnamefont{J.~B.} \bibnamefont{Hastings}},
  \bibinfo{author}{\bibfnamefont{C.}~\bibnamefont{Vettier}}, \bibnamefont{and}
  \bibinfo{author}{\bibfnamefont{O.}~\bibnamefont{Vogt}},
  \bibinfo{journal}{Phys. Rev. Lett.} \textbf{\bibinfo{volume}{62}},
  \bibinfo{pages}{1671} (\bibinfo{year}{1989}).

\bibitem{mgb99}
\bibinfo{author}{\bibfnamefont{D.~F.} \bibnamefont{McMorrow}},
  \bibinfo{author}{\bibfnamefont{D.}~\bibnamefont{Gibbs}}, \bibnamefont{and}
  \bibinfo{author}{\bibfnamefont{J.}~\bibnamefont{Bohr}}, in
  \emph{\bibinfo{booktitle}{Handbook of Physics and Chemistry of Rare Earths}},
  edited by \bibinfo{editor}{\bibnamefont{K.~A.~Gschneidner,~Jr.}}
  \bibnamefont{and} \bibinfo{editor}{\bibfnamefont{L.}~\bibnamefont{Eyring}}
  (\bibinfo{publisher}{Elsevier}, \bibinfo{address}{Amsterdam},
  \bibinfo{year}{1999}), vol.~\bibinfo{volume}{26}, p.~\bibinfo{pages}{1}.

\bibitem{mll99}
\bibinfo{author}{\bibfnamefont{D.}~\bibnamefont{Mannix}},
  \bibinfo{author}{\bibfnamefont{S.}~\bibnamefont{Langridge}},
  \bibinfo{author}{\bibfnamefont{G.~H.} \bibnamefont{Lander}},
  \bibinfo{author}{\bibfnamefont{J.}~\bibnamefont{Rebizant}},
  \bibinfo{author}{\bibfnamefont{M.~J.} \bibnamefont{Longfield}},
  \bibinfo{author}{\bibfnamefont{W.~G.} \bibnamefont{Stirling}},
  \bibinfo{author}{\bibfnamefont{W.~J.} \bibnamefont{Nuttall}},
  \bibinfo{author}{\bibfnamefont{S.}~\bibnamefont{Coburn}},
  \bibinfo{author}{\bibfnamefont{S.}~\bibnamefont{Wasserman}},
  \bibnamefont{and}
  \bibinfo{author}{\bibfnamefont{L.}~\bibnamefont{Soderholm}},
  \bibinfo{journal}{Physica B} \textbf{\bibinfo{volume}{262}},
  \bibinfo{pages}{125} (\bibinfo{year}{1999}).

\bibitem{cck94}
\bibinfo{author}{\bibfnamefont{C.-C.} \bibnamefont{Kao}},
  \bibinfo{author}{\bibfnamefont{C.~T.} \bibnamefont{Chen}},
  \bibinfo{author}{\bibfnamefont{E.~D.} \bibnamefont{Johnson}},
  \bibinfo{author}{\bibfnamefont{J.~B.} \bibnamefont{Hastings}},
  \bibinfo{author}{\bibfnamefont{H.~J.} \bibnamefont{Lin}},
  \bibinfo{author}{\bibfnamefont{G.~H.} \bibnamefont{Ho}},
  \bibinfo{author}{\bibfnamefont{G.}~\bibnamefont{Meigs}},
  \bibinfo{author}{\bibfnamefont{J.~M.} \bibnamefont{Brot}},
  \bibinfo{author}{\bibfnamefont{S.~L.} \bibnamefont{Hulbert}},
  \bibinfo{author}{\bibfnamefont{Y.~U.} \bibnamefont{Idzerda}},
  \bibnamefont{and} \bibinfo{author}{\bibfnamefont{C.}~\bibnamefont{Vettier}},
  \bibinfo{journal}{Phys. Rev. B} \textbf{\bibinfo{volume}{50}},
  \bibinfo{pages}{9599} (\bibinfo{year}{1994}).

\bibitem{tonnerre95}
\bibinfo{author}{\bibfnamefont{J.~M.} \bibnamefont{Tonnerre}},
  \bibinfo{author}{\bibfnamefont{L.}~\bibnamefont{S\`eve}},
  \bibinfo{author}{\bibfnamefont{D.}~\bibnamefont{Raoux}},
  \bibinfo{author}{\bibfnamefont{G.}~\bibnamefont{Soulli\'e}},
  \bibinfo{author}{\bibfnamefont{B.}~\bibnamefont{Rodmacq}}, \bibnamefont{and}
  \bibinfo{author}{\bibfnamefont{P.}~\bibnamefont{Wolfers}},
  \bibinfo{journal}{Phys. Rev. Lett.} \textbf{\bibinfo{volume}{75}},
  \bibinfo{pages}{740} (\bibinfo{year}{1995}).

\bibitem{hks95}
\bibinfo{author}{\bibfnamefont{F.~U.} \bibnamefont{Hillebrecht}},
  \bibinfo{author}{\bibfnamefont{T.}~\bibnamefont{Kinoshita}},
  \bibinfo{author}{\bibfnamefont{D.}~\bibnamefont{Spanke}},
  \bibinfo{author}{\bibfnamefont{J.}~\bibnamefont{Dresselhaus}},
  \bibinfo{author}{\bibfnamefont{C.}~\bibnamefont{Roth}},
  \bibinfo{author}{\bibfnamefont{H.~B.} \bibnamefont{Rose}}, \bibnamefont{and}
  \bibinfo{author}{\bibfnamefont{E.}~\bibnamefont{Kisker}},
  \bibinfo{journal}{Phys. Rev. Lett.} \textbf{\bibinfo{volume}{75}},
  \bibinfo{pages}{2224} (\bibinfo{year}{1995}).

\bibitem{cic98}
\bibinfo{author}{\bibfnamefont{V.}~\bibnamefont{Chakarian}},
  \bibinfo{author}{\bibfnamefont{Y.~U.} \bibnamefont{Idzerda}},
  \bibnamefont{and} \bibinfo{author}{\bibfnamefont{C.~T.} \bibnamefont{Chen}},
  \bibinfo{journal}{Phys. Rev. B} \textbf{\bibinfo{volume}{57}},
  \bibinfo{pages}{5312} (\bibinfo{year}{1998}).

\bibitem{shp98}
\bibinfo{author}{\bibfnamefont{M.}~\bibnamefont{Sacchi}},
  \bibinfo{author}{\bibfnamefont{C.~F.} \bibnamefont{Hague}},
  \bibinfo{author}{\bibfnamefont{L.}~\bibnamefont{Pasquali}},
  \bibinfo{author}{\bibfnamefont{A.}~\bibnamefont{Mirone}},
  \bibinfo{author}{\bibfnamefont{J.-M.} \bibnamefont{Mariot}},
  \bibinfo{author}{\bibfnamefont{P.}~\bibnamefont{Isberg}},
  \bibinfo{author}{\bibfnamefont{E.~M.} \bibnamefont{Gullikson}},
  \bibnamefont{and} \bibinfo{author}{\bibfnamefont{J.~H.}
  \bibnamefont{Underwood}}, \bibinfo{journal}{Phys. Rev. Lett.}
  \textbf{\bibinfo{volume}{81}}, \bibinfo{pages}{1521} (\bibinfo{year}{1998}).

\bibitem{kok00}
\bibinfo{author}{\bibfnamefont{J.~B.} \bibnamefont{Kortright}}
  \bibnamefont{and} \bibinfo{author}{\bibfnamefont{S.-K.} \bibnamefont{Kim}},
  \bibinfo{journal}{Phys. Rev. B} \textbf{\bibinfo{volume}{62}},
  \bibinfo{pages}{12216} (\bibinfo{year}{2000}).

\bibitem{schuessl01}
\bibinfo{author}{\bibfnamefont{C.}~\bibnamefont{Sch\"u\ss{}ler-Langeheine}},
  \bibinfo{author}{\bibfnamefont{E.}~\bibnamefont{Weschke}},
  \bibinfo{author}{\bibfnamefont{A.~Y.} \bibnamefont{Grigoriev}},
  \bibinfo{author}{\bibfnamefont{H.}~\bibnamefont{Ott}},
  \bibinfo{author}{\bibfnamefont{R.}~\bibnamefont{Meier}},
  \bibinfo{author}{\bibfnamefont{D.~V.} \bibnamefont{Vyalikh}},
  \bibinfo{author}{\bibfnamefont{C.}~\bibnamefont{Mazumdar}},
  \bibinfo{author}{\bibfnamefont{C.}~\bibnamefont{Sutter}},
  \bibinfo{author}{\bibfnamefont{D.}~\bibnamefont{Abernathy}},
  \bibinfo{author}{\bibfnamefont{G.}~\bibnamefont{Gr\"ubel}}, \bibnamefont{and}
  \bibinfo{author}{\bibfnamefont{G.}~\bibnamefont{Kaindl}},
  \bibinfo{journal}{J. Electron Spectrosc. Relat. Phenomen.}
  \textbf{\bibinfo{volume}{114-116}}, \bibinfo{pages}{953}
  (\bibinfo{year}{2001}).

\bibitem{mog01}
\bibinfo{author}{\bibfnamefont{H.-C.} \bibnamefont{Mertins}},
  \bibinfo{author}{\bibfnamefont{P.~M.} \bibnamefont{Oppeneer}},
  \bibinfo{author}{\bibfnamefont{J.}~\bibnamefont{Kune\v{s}}},
  \bibinfo{author}{\bibfnamefont{A.}~\bibnamefont{Gaupp}},
  \bibinfo{author}{\bibfnamefont{D.}~\bibnamefont{Abramsohn}},
  \bibnamefont{and}
  \bibinfo{author}{\bibfnamefont{F.}~\bibnamefont{Sch\"afers}},
  \bibinfo{journal}{Phys. Rev. Lett.} \textbf{\bibinfo{volume}{87}},
  \bibinfo{pages}{47401} (\bibinfo{year}{2001}).

\bibitem{tsb98}
\bibinfo{author}{\bibfnamefont{J.~M.} \bibnamefont{Tonnerre}},
  \bibinfo{author}{\bibfnamefont{L.}~\bibnamefont{Seve}},
  \bibinfo{author}{\bibfnamefont{A.}~\bibnamefont{Barbara-Dechelette}},
  \bibinfo{author}{\bibfnamefont{F.}~\bibnamefont{Bartolome}},
  \bibinfo{author}{\bibfnamefont{D.}~\bibnamefont{Raoux}},
  \bibinfo{author}{\bibfnamefont{V.}~\bibnamefont{Chakarian}},
  \bibinfo{author}{\bibfnamefont{C.~C.} \bibnamefont{Kao}},
  \bibinfo{author}{\bibfnamefont{H.}~\bibnamefont{Fischer}},
  \bibinfo{author}{\bibfnamefont{S.}~\bibnamefont{Andrieu}}, \bibnamefont{and}
  \bibinfo{author}{\bibfnamefont{O.}~\bibnamefont{Fruchart}},
  \bibinfo{journal}{J. Appl. Phys.} \textbf{\bibinfo{volume}{83}},
  \bibinfo{pages}{6293} (\bibinfo{year}{1998}).

\bibitem{icf99}
\bibinfo{author}{\bibfnamefont{Y.~U.} \bibnamefont{Idzerda}},
  \bibinfo{author}{\bibfnamefont{V.}~\bibnamefont{Chakarian}},
  \bibnamefont{and} \bibinfo{author}{\bibfnamefont{J.~W.}
  \bibnamefont{Freeland}}, \bibinfo{journal}{Phys. Rev. Lett.}
  \textbf{\bibinfo{volume}{82}}, \bibinfo{pages}{1562} (\bibinfo{year}{1999}).

\bibitem{wbh99}
\bibinfo{author}{\bibfnamefont{N.}~\bibnamefont{Weber}},
  \bibinfo{author}{\bibfnamefont{C.}~\bibnamefont{Bethke}}, \bibnamefont{and}
  \bibinfo{author}{\bibfnamefont{F.~U.} \bibnamefont{Hillebrecht}},
  \bibinfo{journal}{J. Appl. Phys.} \textbf{\bibinfo{volume}{85}},
  \bibinfo{pages}{4946} (\bibinfo{year}{1999}).

\bibitem{kko00}
\bibinfo{author}{\bibfnamefont{J.~B.} \bibnamefont{Kortright}},
  \bibinfo{author}{\bibfnamefont{S.-K.} \bibnamefont{Kim}}, \bibnamefont{and}
  \bibinfo{author}{\bibfnamefont{H.}~\bibnamefont{Ohldag}},
  \bibinfo{journal}{Phys. Rev. B} \textbf{\bibinfo{volume}{61}},
  \bibinfo{pages}{64} (\bibinfo{year}{2000}).

\bibitem{hkt00}
\bibinfo{author}{\bibfnamefont{O.}~\bibnamefont{Hellwig}},
  \bibinfo{author}{\bibfnamefont{J.~B.} \bibnamefont{Kortright}},
  \bibinfo{author}{\bibfnamefont{K.}~\bibnamefont{Takano}}, \bibnamefont{and}
  \bibinfo{author}{\bibfnamefont{E.~E.} \bibnamefont{Fullerton}},
  \bibinfo{journal}{Phys. Rev. B} \textbf{\bibinfo{volume}{62}},
  \bibinfo{pages}{11694} (\bibinfo{year}{2000}).

\bibitem{ggj01}
\bibinfo{author}{\bibfnamefont{J.}~\bibnamefont{Geissler}},
  \bibinfo{author}{\bibfnamefont{E.}~\bibnamefont{Goering}},
  \bibinfo{author}{\bibfnamefont{M.}~\bibnamefont{Justen}},
  \bibinfo{author}{\bibfnamefont{F.}~\bibnamefont{Weigand}},
  \bibinfo{author}{\bibfnamefont{G.}~\bibnamefont{Sch\"utz}},
  \bibinfo{author}{\bibfnamefont{J.}~\bibnamefont{Langer}},
  \bibinfo{author}{\bibfnamefont{D.}~\bibnamefont{Schmitz}},
  \bibinfo{author}{\bibfnamefont{H.}~\bibnamefont{Maletta}}, \bibnamefont{and}
  \bibinfo{author}{\bibfnamefont{R.}~\bibnamefont{Mattheis}},
  \bibinfo{journal}{Phys. Rev. B} \textbf{\bibinfo{volume}{65}},
  \bibinfo{pages}{020405} (\bibinfo{year}{2001}).

\bibitem{sts00}
\bibinfo{author}{\bibfnamefont{S.~A.} \bibnamefont{Stepanov}} \bibnamefont{and}
  \bibinfo{author}{\bibfnamefont{S.~K.} \bibnamefont{Sinha}},
  \bibinfo{journal}{Phys. Rev. B} \textbf{\bibinfo{volume}{61}},
  \bibinfo{pages}{15302} (\bibinfo{year}{2000}).

\bibitem{fjs98}
\bibinfo{author}{\bibfnamefont{E.~E.} \bibnamefont{Fullerton}},
  \bibinfo{author}{\bibfnamefont{J.~S.} \bibnamefont{Jiang}},
  \bibinfo{author}{\bibfnamefont{C.~H.} \bibnamefont{Sowers}},
  \bibinfo{author}{\bibfnamefont{J.~E.} \bibnamefont{Pearson}},
  \bibnamefont{and} \bibinfo{author}{\bibfnamefont{S.~D.} \bibnamefont{Bader}},
  \bibinfo{journal}{Appl. Phys. Lett.} \textbf{\bibinfo{volume}{72}},
  \bibinfo{pages}{380} (\bibinfo{year}{1998}).

\bibitem{nak99}
\bibinfo{author}{\bibfnamefont{Y.}~\bibnamefont{Nakamura}},
  \bibinfo{journal}{J. Magn. Magn. Mater.} \textbf{\bibinfo{volume}{200}},
  \bibinfo{pages}{634} (\bibinfo{year}{1999}).

\bibitem{shv01}
\bibinfo{author}{\bibfnamefont{K.}~\bibnamefont{Starke}},
  \bibinfo{author}{\bibfnamefont{F.}~\bibnamefont{Heigl}},
  \bibinfo{author}{\bibfnamefont{A.}~\bibnamefont{Vollmer}},
  \bibinfo{author}{\bibfnamefont{M.}~\bibnamefont{Weiss}},
  \bibinfo{author}{\bibfnamefont{G.}~\bibnamefont{Reichardt}},
  \bibnamefont{and} \bibinfo{author}{\bibfnamefont{G.}~\bibnamefont{Kaindl}},
  \bibinfo{journal}{Phys. Rev. Lett.} \textbf{\bibinfo{volume}{86}},
  \bibinfo{pages}{3415} (\bibinfo{year}{2001}).

\bibitem{UE56}
\bibinfo{author}{\bibfnamefont{M.~R.} \bibnamefont{Weiss}},
  \bibinfo{author}{\bibfnamefont{R.}~\bibnamefont{Follath}},
  \bibinfo{author}{\bibfnamefont{K.~J.~S.} \bibnamefont{Sawhney}},
  \bibinfo{author}{\bibfnamefont{F.}~\bibnamefont{Senf}},
  \bibinfo{author}{\bibfnamefont{J.}~\bibnamefont{Bahrdt}},
  \bibinfo{author}{\bibfnamefont{W.}~\bibnamefont{Frentrup}},
  \bibinfo{author}{\bibfnamefont{A.}~\bibnamefont{Gaupp}},
  \bibinfo{author}{\bibfnamefont{S.}~\bibnamefont{Sasaki}},
  \bibinfo{author}{\bibfnamefont{M.}~\bibnamefont{Scheer}},
  \bibinfo{author}{\bibfnamefont{H.-C.} \bibnamefont{Mertins}},
  \bibinfo{author}{\bibfnamefont{D.}~\bibnamefont{Abramsohn}},
  \bibinfo{author}{\bibfnamefont{F.}~\bibnamefont{Sch\"afers}}, 
  \bibinfo{author}{\bibfnamefont{W.}~\bibnamefont{Kuch}},
  \bibnamefont{and} \bibinfo{author}{\bibfnamefont{W.}~\bibnamefont{Mahler}},
  \bibinfo{journal}{Nucl. Instrum. Meth. Phys. Res. A}
  \textbf{\bibinfo{volume}{467}}, \bibinfo{pages}{449} (\bibinfo{year}{2001}).

\bibitem{sna97}
\bibinfo{author}{\bibfnamefont{K.}~\bibnamefont{Starke}},
  \bibinfo{author}{\bibfnamefont{E.}~\bibnamefont{Navas}},
  \bibinfo{author}{\bibfnamefont{E.}~\bibnamefont{Arenholz}},
  \bibinfo{author}{\bibfnamefont{Z.}~\bibnamefont{Hu}},
  \bibinfo{author}{\bibfnamefont{L.}~\bibnamefont{Baumgarten}},
  \bibinfo{author}{\bibfnamefont{G.}~\bibnamefont{van~der Laan}},
  \bibinfo{author}{\bibfnamefont{C.-T.} \bibnamefont{Chen}}, \bibnamefont{and}
  \bibinfo{author}{\bibfnamefont{G.}~\bibnamefont{Kaindl}},
  \bibinfo{journal}{Phys. Rev. B} \textbf{\bibinfo{volume}{55}},
  \bibinfo{pages}{2672} (\bibinfo{year}{1997}).

\bibitem{ahs97}
\bibinfo{author}{\bibfnamefont{D.}~\bibnamefont{Alders}},
  \bibinfo{author}{\bibfnamefont{T.}~\bibnamefont{Hibma}},
  \bibinfo{author}{\bibfnamefont{G.~A.} \bibnamefont{Sawatzky}},
  \bibinfo{author}{\bibfnamefont{K.~C.} \bibnamefont{Cheung}},
  \bibinfo{author}{\bibfnamefont{G.~E.} \bibnamefont{van Dorssen}},
  \bibinfo{author}{\bibfnamefont{M.~D.} \bibnamefont{Roper}},
  \bibinfo{author}{\bibfnamefont{H.~A.} \bibnamefont{Padmore}},
  \bibinfo{author}{\bibfnamefont{G.}~\bibnamefont{van~der Laan}},
  \bibinfo{author}{\bibfnamefont{J.}~\bibnamefont{Vogel}}, \bibnamefont{and}
  \bibinfo{author}{\bibfnamefont{M.}~\bibnamefont{Sacchi}},
  \bibinfo{journal}{J. Appl. Phys.} \textbf{\bibinfo{volume}{82}},
  \bibinfo{pages}{3120} (\bibinfo{year}{1997}).

\bibitem{sta00}
\bibinfo{author}{\bibfnamefont{K.}~\bibnamefont{Starke}},
  \emph{\bibinfo{title}{Magnetic Dichroism in Core-Level Photoemission}}
  (\bibinfo{publisher}{Springer}, \bibinfo{address}{Berlin},
  \bibinfo{year}{2000}).

\bibitem{magnet}
\bibinfo{author}{\bibfnamefont{F.}~\bibnamefont{Heigl}},
  \bibinfo{author}{\bibfnamefont{O.}~\bibnamefont{Krupin}},
  \bibinfo{author}{\bibfnamefont{G.}~\bibnamefont{Kaindl}}, \bibnamefont{and}
  \bibinfo{author}{\bibfnamefont{K.}~\bibnamefont{Starke}},
  \bibinfo{journal}{Rev. Sci. Instrum.} \textbf{\bibinfo{volume}{73}},
  \bibinfo{pages}{369} (\bibinfo{year}{2002}).

\bibitem{sed92}
\bibinfo{author}{\bibfnamefont{K.}~\bibnamefont{Starke}},
  \bibinfo{author}{\bibfnamefont{K.}~\bibnamefont{Ertl}}, \bibnamefont{and}
  \bibinfo{author}{\bibfnamefont{V.}~\bibnamefont{Dose}},
  \bibinfo{journal}{Phys. Rev. B} \textbf{\bibinfo{volume}{46}},
  \bibinfo{pages}{9709} (\bibinfo{year}{1992}).

\bibitem{vt88}
\bibinfo{author}{\bibfnamefont{G.}~\bibnamefont{van~der Laan}}
  \bibnamefont{and} \bibinfo{author}{\bibfnamefont{B.~T.} \bibnamefont{Thole}},
  \bibinfo{journal}{J. Electron Spectrosc. Relat. Phenomen.}
  \textbf{\bibinfo{volume}{46}}, \bibinfo{pages}{123} (\bibinfo{year}{1988}).

\bibitem{nsi99}
\bibinfo{author}{\bibfnamefont{R.}~\bibnamefont{Nakajima}},
  \bibinfo{author}{\bibfnamefont{J.}~\bibnamefont{St\"ohr}}, \bibnamefont{and}
  \bibinfo{author}{\bibfnamefont{Y.~U.} \bibnamefont{Idzerda}},
  \bibinfo{journal}{Phys. Rev. B} \textbf{\bibinfo{volume}{59}},
  \bibinfo{pages}{6421} (\bibinfo{year}{1999}).

\bibitem{muto94}
\bibinfo{author}{\bibfnamefont{S.}~\bibnamefont{Muto}},
  \bibinfo{author}{\bibfnamefont{S.-Y.} \bibnamefont{Park}},
  \bibinfo{author}{\bibfnamefont{S.}~\bibnamefont{Imada}},
  \bibinfo{author}{\bibfnamefont{K.}~\bibnamefont{Yamaguchi}},
  \bibinfo{author}{\bibfnamefont{Y.}~\bibnamefont{Kagoshima}},
  \bibnamefont{and} \bibinfo{author}{\bibfnamefont{T.}~\bibnamefont{Miyhara}},
  \bibinfo{journal}{J. Phys. Soc. Jpn.} \textbf{\bibinfo{volume}{63}},
  \bibinfo{pages}{1179} (\bibinfo{year}{1994}).

\bibitem{henke}
\bibinfo{author}{\bibfnamefont{B.~L.} \bibnamefont{Henke}},
  \bibinfo{author}{\bibfnamefont{E.~M.} \bibnamefont{Gullikson}},
  \bibnamefont{and} \bibinfo{author}{\bibfnamefont{J.~C.} \bibnamefont{Davis}},
  \bibinfo{journal}{At. Data Nucl. Data Tables} \textbf{\bibinfo{volume}{54}},
  \bibinfo{pages}{180} (\bibinfo{year}{1993}), \bibinfo{note}{www-cxro. lbl.
  gov/optical\_constants}.

\bibitem{zfg67}
\bibinfo{author}{\bibfnamefont{T.~M.} \bibnamefont{Zimkina}},
  \bibinfo{author}{\bibfnamefont{V.~A.} \bibnamefont{Fomichev}},
  \bibinfo{author}{\bibfnamefont{S.~A.} \bibnamefont{Gribivskii}},
  \bibnamefont{and} \bibinfo{author}{\bibfnamefont{I.~I.}
  \bibnamefont{Zhukova}}, \bibinfo{journal}{Fiz. Tverd. Tela (Leningrad)}
  \textbf{\bibinfo{volume}{9}}, \bibinfo{pages}{1147} (\bibinfo{year}{1967}),
  \bibinfo{note}{[Sov. Phys. Solid State {\bf 9}, 1128 (1967)]}.

\bibitem{rmp89}
\bibinfo{author}{\bibfnamefont{M.}~\bibnamefont{Richter}},
  \bibinfo{author}{\bibfnamefont{M.}~\bibnamefont{Meyer}},
  \bibinfo{author}{\bibfnamefont{M.}~\bibnamefont{Pahler}},
  \bibinfo{author}{\bibfnamefont{T.}~\bibnamefont{Prescher}},
  \bibinfo{author}{\bibfnamefont{E.}~\bibnamefont{v.~Raven}},
  \bibinfo{author}{\bibfnamefont{B.}~\bibnamefont{Sonntag}}, \bibnamefont{and}
  \bibinfo{author}{\bibfnamefont{H.-E.} \bibnamefont{Wetzel}},
  \bibinfo{journal}{Phys. Rev. A} \textbf{\bibinfo{volume}{40}},
  \bibinfo{pages}{7007} (\bibinfo{year}{1989}).

\bibitem{pva99}
\bibinfo{author}{\bibfnamefont{K.-E.} \bibnamefont{Peiponen}},
  \bibinfo{author}{\bibfnamefont{E.~M.} \bibnamefont{Vartiainen}},
  \bibnamefont{and} \bibinfo{author}{\bibfnamefont{T.}~\bibnamefont{Asakura}},
  \emph{\bibinfo{title}{Dispersion, Complex Analysis and Optical Spectroscopy}}
  (\bibinfo{publisher}{Springer}, \bibinfo{address}{Berlin},
  \bibinfo{year}{1999}).

\bibitem{zvk97}
\bibinfo{author}{\bibfnamefont{A.~K.} \bibnamefont{Zvezdin}} \bibnamefont{and}
  \bibinfo{author}{\bibfnamefont{V.~A.} \bibnamefont{Kotov}},
  \emph{\bibinfo{title}{Modern Magnetooptics and Magnetooptical Materials}}
  (\bibinfo{publisher}{Institute of Physics Publishing},
  \bibinfo{address}{Bristol}, \bibinfo{year}{1997}).

\end{thebibliography}

\end{document}